\documentclass[aps,prl,singlecolumn,superscriptaddress,nofootinbib,floatfix]{revtex4-2}
\usepackage{amsmath}
\usepackage{graphicx}
\usepackage{dcolumn}
\usepackage{bm}
\usepackage{amssymb}
\usepackage{dsfont}
\usepackage{url}
\usepackage{caption,subcaption}
\usepackage[colorlinks=true,allcolors=blue]{hyperref}

\newcommand{\supplemental}[1]{{\vspace{10 pt}\centerline{\textbf{#1}}}}

\begin{document}

\title{Exact results for the Boltzmann collision operator in $\lambda\phi^4$ theory}
\author{Gabriel S.~Denicol}
\email{gsdenicol@id.uff.br}
\affiliation{Instituto de F\'{\i}sica, Universidade Federal Fluminense, Niter\'{o}i, Rio
de Janeiro, Brazil}
\author{Jorge Noronha} 
\email{jn0508@illinois.edu}
\affiliation{Illinois Center for Advanced Studies of the Universe\\ Department of Physics, 
University of Illinois at Urbana-Champaign, Urbana, IL 61801, USA}

\begin{abstract}
We \emph{analytically} determine all the eigenvalues and eigenfunctions of the linearized Boltzmann collision operator in massless scalar $\lambda \phi^4$ theory in the high-temperature (classical) regime. This is used to exactly compute the shear viscosity and particle diffusion transport coefficients of this system. The corresponding relaxation time approximation for this linearized Boltzmann equation is also derived.   
\end{abstract}

\maketitle

\noindent
{\it \textbf{Introduction}} -- The relativistic Boltzmann equation is a key tool in the description of the complex behavior displayed by dilute relativistic gases in many fields. Applications can be found in many areas, such as high-energy nuclear physics and the quark-gluon plasma \cite{Heinz:1984yq, Bass:1998ca,Arnold:2000dr,Xu:2004mz,Denicol:2012cn,Weil:2016zrk}, relativistic astrophysical plasmas \cite{Most:2021uck}, neutrino transport in supernova \cite{Janka:2012wk,RevModPhys.85.245}, and cosmology \cite{Ma:1995ey,Weinberg:2008zzc}. This transport equation can be derived as an approximation of the weak-coupling limit of quantum field theories, such as for instance self-interacting $\lambda \phi^4$ scalar field theory \cite{Jeon:1994if}, or more complicated systems including fermions and gauge fields \cite{Arnold:2002zm}. Besides general statements concerning conservation laws and entropy production \cite{degroot}, most of what is known about the Boltzmann equation, both in the relativistic and non-relativistic regimes, has been obtained through numerical simulations. Analytical solutions are extremely rare and have been determined only for isotropic, homogeneous systems and, even so, for simple interactions   \cite{KrookWu1976,KrookWu1977,Bobylev,Bazow:2015dha,Bazow:2016oky}. 

The linearized collision operator, which determines the hydrodynamic properties of the system such as its transport coefficients \cite{ChapmanCowling}, plays a central role in our understanding of gases. In the non-relativistic regime, only a handful of results for this operator have been determined analytically. Remarkably,  the eigenfunctions of the linearized collision term can be written in terms of Hermite polynomials for the case of Maxwell molecules (i.e., particles interacting via a $\sim 1/r^4$ potential), as shown by Wang Chang and Uhlenbeck \cite{ChangUhlenbeck,cercignani:90mathematical}. The eigenvalues of this operator were also determined for these interactions, but not in a simple analytical form. In the relativistic regime, no analytical results for the eigenvalues and/or the eigenfunctions of the collision operator are known for any type of interactions.

In this letter we fill in this gap and \emph{analytically} obtain the full set of eigenvalues and eigenfunctions of the linearized Boltzmann collision operator in massless scalar $\lambda \phi^4$ theory in the high-temperature (classical) regime. We show that the eigenvalue spectrum is discrete and bounded from below and the eigenfunctions are exactly defined in terms of Laguerre polynomials and a covariant generalization of spherical harmonics \cite{1974JMP....15.1116A}. This allows us to compute the shear and particle diffusion transport coefficients for this system in exact form for the first time. Knowledge of the exact eigenvalues and eigenvectors also leads to the determination of the correct relaxation time approximation for this theory, which is done for the first time in this work.   

We use the mostly minus convention for the Minkowski
metric, $g_{\mu\nu}=\textrm{diag}(1,-1,-1,-1)$, and natural units, i.e. $\hbar=c=k_{B}=1$. Throughout the text, the spacetime dependence of some
functions is omitted and the momentum dependence is denoted by a sub-index,
so that $f(x,p)=f_{\mathbf{p}}$.

\noindent
{\it \textbf{Linearized Boltzmann equation}} -- The relativistic Boltzmann equation is an integro-differential equation for the single-particle momentum distribution
function $f_{\mathbf{p}}$. If the system is close to thermodynamic equilibrium, it may be effectively described by the linearized Boltzmann equation \cite{degroot,Denicol:2021},
\begin{equation}
k^{\mu }\partial _{\mu }f_{\mathbf{k}}=f_{0\mathbf{k}}\hat{L}\phi _{\mathbf{k%
}},
\end{equation}%
where $f_{0\mathbf{k}}=\exp \left( \alpha -\beta E_{\mathbf{k}}\right) $
is the equilibrium distribution function (classical statistics), with $\alpha $ being the
thermal potential, $\beta =1/T$ the inverse temperature, and $E_{\mathbf{k}}=u_{\mu}k^\mu$ with $u^{\mu}$ being a
unitary time-like 4-vector (usually identified as the fluid's 4-velocity). For the sake
of convenience, we further introduced the field $\phi _{\mathbf{k}}=\left(
f_{\mathbf{k}}-f_{0\mathbf{k}}\right) /f_{0\mathbf{k}}$, which quantifies
the deviation of the system from equilibrium, and the linearized
collision operator, $\hat{L}$.

We consider a gas of massless scalar field particles with quartic (tree-level) self-interactions ($\lambda \varphi ^{4}$) in the classical limit. In this case, the collision operator takes the following form \cite{Denicol:2021}
\begin{equation}
\hat{L}\phi _{\mathbf{k}}=\frac{g}{2}\int dK^{\prime }dPdP^{\prime }f_{0%
\mathbf{k}^{\prime }}\left( 2\pi \right) ^{5}\delta ^{(4)}\left( k+k^{\prime
}-p-p^{\prime }\right) \left( \phi _{\mathbf{p}}+\phi _{\mathbf{p}^{\prime
}}-\phi _{\mathbf{k}}-\phi _{\mathbf{k}^{\prime }}\right) .
\end{equation}%
Here, we defined the Lorentz invariant momentum-space volume, $dK\equiv d^{3}k/
[ \left( 2\pi \right) ^{3}k^{0}] $, and used that the total
cross-section is given by $\sigma_{T}=g/s$,
with $s$ being the Mandelstam variable, $s=\left( k^{\mu }+k^{\prime \mu
}\right) \left( k_{\mu }+k_{\mu }^{\prime }\right) $.
The factor $g$ is a constant and quantifies the strength of the interaction
-- in terms of the coupling constant of the scalar theory, $\lambda $, it is
given by $g=\lambda ^{2}/\left( 32\pi \right) $. It is convenient to separate the linear operator $\hat{L}$ into ``gain" and ``loss" contributions, in the following way
\begin{equation}
\hat{L}\equiv \hat{K}_{\mathrm{gain}}-\hat{K}_{\mathrm{loss}},
\end{equation}
where the new operators are defined as 
\begin{eqnarray}
\hat{K}_{\mathrm{gain}}\phi _{\mathbf{k}} &=&g\int dK^{\prime }dPdP^{\prime
}f_{0\mathbf{k}^{\prime }}\left( 2\pi \right) ^{5}\delta ^{4}\left(
k+k^{\prime }-p-p^{\prime }\right) \phi _{\mathbf{p}}, \\
\hat{K}_{\mathrm{loss}}\phi _{\mathbf{k}} &=&\frac{g}{2}\int dK^{\prime
}dPdP^{\prime }f_{0\mathbf{k}^{\prime }}\left( 2\pi \right) ^{5}\delta
^{4}\left( k+k^{\prime }-p-p^{\prime }\right) \left( \phi _{\mathbf{k}}+\phi
_{\mathbf{k}^{\prime }}\right) .
\end{eqnarray}

\noindent
{\it \textbf{Spectral decomposition}} -- The goal of this letter is to demonstrate that the tensors $L_{n}^{\left(
2\ell +1\right) }\left( \beta E_{\mathbf{k}}\right) k^{\left\langle \mu
_{1}\right. }\ldots k^{\left. \mu _{\ell }\right\rangle }$ are the complete set of
eigenfunctions of $\hat{L}$, with eigenvalues that will also be calculated in this work. Here, $L_{n}^{\left( 2\ell +1\right) }\left( \beta E_{\mathbf{k%
}}\right) \equiv L_{n\mathbf{k}}^{\left( 2\ell +1\right) }$ corresponds to
the associated Laguerre polynomial of degree $n$ and $k^{\left\langle \mu
_{1}\right. }\ldots k^{\left. \mu _{\ell }\right\rangle }$ are irreducible
(with respect to the Lorentz little group) tensors of rank $\ell$ \cite{degroot} constructed from the 4-momentum, $k^{\mu }$. Irreducible tensors are defined by using the symmetrized and, for $m>1$, traceless, projections
orthogonal to $u^{\mu }$ as,%
\begin{equation}
A^{\left\langle \mu _{1}\right. \cdots \left. \mu _{\ell }\right\rangle
}=\Delta _{\nu _{1}\cdots \nu _{\ell }}^{\mu _{1}\cdots \mu _{\ell }}A^{\nu
_{1}\cdots \nu _{\ell }},
\end{equation}%
where $\Delta _{\nu _{1}\cdots \nu _{\ell }}^{\mu _{1}\cdots \mu _{\ell }}$
are traceless and symmetric projection operators constructed from $\Delta
_{\nu }^{\mu }=g_{\nu }^{\mu }-u^{\mu }u_{\nu }$ \cite{Denicol:2021}. We note that both $%
L_{n\mathbf{k}}^{\left( 2\ell +1\right) }$ and $k^{\left\langle \mu
_{1}\right. }\ldots k^{\left. \mu _{\ell }\right\rangle }$ form a complete
and orthogonal basis \cite{ANDERSON1974466,degroot,AbramowitzStegun}, with $k^{\left\langle \mu _{1}\right. }\ldots
k^{\left. \mu _{\ell }\right\rangle }$ being equivalent to spherical
harmonics in the local rest frame of the fluid \cite{1974JMP....15.1116A}.

We shall demonstrate that $L_{n\mathbf{k}}^{\left( 2\ell +1\right)
}k^{\left\langle \mu _{1}\right. }\ldots k^{\left. \mu _{\ell }\right\rangle
}$ are eigenfunctions of $\hat{L}$ by directly applying the
operator on this function%
\begin{equation}
\hat{L}L_{n\mathbf{k}}^{\left( 2\ell +1\right) }k^{\left\langle \mu
_{1}\right. }\ldots k^{\left. \mu _{\ell }\right\rangle }=\hat{K}_{\mathrm{%
gain}}L_{n\mathbf{k}}^{\left( 2\ell +1\right) }k^{\left\langle \mu
_{1}\right. }\ldots k^{\left. \mu _{\ell }\right\rangle }-\hat{K}_{\mathrm{%
loss}}L_{n\mathbf{k}}^{\left( 2\ell +1\right) }k^{\left\langle \mu
_{1}\right. }\ldots k^{\left. \mu _{\ell }\right\rangle }.
\end{equation}%
The second contribution, related to the loss term, can be calculated easily%
\begin{equation}
\label{loss}
\hat{K}_{\mathrm{loss}}L_{n\mathbf{k}}^{\left( 2\ell +1\right)
}k^{\left\langle \mu _{1}\right. }\ldots k^{\left. \mu _{\ell }\right\rangle
}=g\mathcal{M}\frac{1+\delta _{\ell 0}\delta _{n0}}{2}L_{n\mathbf{k}
}^{\left( 2\ell +1\right) }k^{\left\langle \mu _{1}\right. }\ldots k^{\left.
\mu _{\ell }\right\rangle },
\end{equation}
where we used the orthogonality relations satisfied by the Laguerre
polynomials and irreducible tensors \cite{Denicol:2021}, and also used the following identity \cite{Bazow:2015dha}
\begin{equation}
\int dPdP^{\prime } \left( 2\pi \right) ^{5}\delta
^{4}\left( k+k^{\prime }-p-p^{\prime }\right) =1.
\end{equation}
We further defined the quantity
\begin{equation}
\int dK f_{0\mathbf{k}}=\frac{\exp \left( \alpha \right) }{
2\pi ^{2}\beta ^{2}}\equiv \mathcal{M}.
\end{equation}

The gain term is more complicated and will be evaluated below. Overall, we have to calculate the integral 
\begin{equation}
\hat{K}_{\mathrm{gain}}L_{n\mathbf{k}}^{\left( 2\ell +1\right)
}k^{\left\langle \mu _{1}\right. }\ldots k^{\left. \mu _{\ell }\right\rangle
}=g\int dK^{\prime }dPdP^{\prime }\left. {}\right. \left( 2\pi \right)
^{5}\delta ^{4}\left( k+k^{\prime }-p-p^{\prime }\right) f_{0\mathbf{k}
^{\prime }}L_{n\mathbf{p}}^{\left( 2\ell +1\right) }p^{\left\langle \mu
_{1}\right. }\ldots p^{\left. \mu _{\ell }\right\rangle }.
\end{equation}
First we note that the projection operator $\Delta _{\nu _{1}\cdots \nu
_{\ell }}^{\mu _{1}\cdots \mu _{\ell }}$ commutes with $\hat{K}_{\mathrm{gain%
}}$ and, consequently, $\hat{K}_{\mathrm{gain}}L_{n\mathbf{k}}^{\left( 2\ell
+1\right) }k^{\left\langle \mu _{1}\right. }\ldots k^{\left. \mu _{\ell
}\right\rangle }$ must also be an irreducible tensor of rank $\ell $. Since
the integral depends only on two external vectors, $u^{\mu }$ and $k^{\mu }$, the resulting integral must be an irreducible tensor constructed solely in terms of these
4--vectors and the metric tensor. Therefore, it must have the following form
\begin{equation}
\hat{K}_{\mathrm{gain}}L_{n\mathbf{k}}^{\left( 2\ell +1\right)
}k^{\left\langle \mu _{1}\right. }\ldots k^{\left. \mu _{\ell }\right\rangle
}=\mathcal{A}_{n\ell }k^{\left\langle \mu _{1}\right. }\ldots k^{\left. \mu
_{\ell }\right\rangle },
\end{equation}
where we introduced the Lorentz scalar integral%
\begin{equation}
\mathcal{A}_{n\ell }=g\frac{\left( 2\ell -1\right) !!}{\left( -1\right)
^{\ell }\ell !E_{\mathbf{k}}^{2\ell }}\int dK^{\prime }f_{0\mathbf{k}
^{\prime }}k_{\left\langle \mu _{1}\right. }\ldots k_{\left. \mu _{\ell
}\right\rangle }\int dPdP^{\prime }\left. {}\right. \left( 2\pi \right)
^{5}\delta ^{4}\left( k+k^{\prime }-p-p^{\prime }\right) L_{n\mathbf{p}
}^{\left( 2\ell +1\right) }p^{\mu _{1}}\ldots p^{\mu _{\ell }},
\end{equation}
and used that, in the massless limit, \cite{Denicol:2021}
\begin{equation}
k^{\left\langle \mu _{1}\right. }\cdots k^{\left. \mu _{\ell }\right\rangle
}k_{\left\langle \mu _{1}\right. }\cdots k_{\left. \mu _{\ell }\right\rangle
}=\frac{\left( -1\right) ^{\ell }\ell !E_{\mathbf{k}}^{2\ell }}{\left( 2\ell
-1\right) !!}.
\end{equation}

The quantity $\mathcal{A}_{n\ell}$ can be evaluated using the generating function
\begin{equation}
G_{t}\equiv \frac{\left( t\beta \right) ^{2}}{\left[ -t\beta \left(
1-t\right) \right] ^{\ell +2}}\int dPdP^{\prime }\left. {}\right. \left(
2\pi \right) ^{5}\delta ^{(4)}\left( k+k^{\prime }-p-p^{\prime }\right) \exp
\left( -\frac{t}{1-t}\beta E_{\mathbf{p}}\right) .
\end{equation}
If we consider $u^{\mu }$ as an arbitrary time-like 4-vector, with magnitude $u^2\equiv u_\alpha u^\alpha$, we obtain the
desired integral by
\begin{equation}
\mathcal{A}_{n\ell }=g\frac{\left( 2\ell -1\right) !!}{\left( -1\right)
^{\ell }\ell !E_{\mathbf{k}}^{2\ell }}\lim_{t\rightarrow 0}\frac{1}{n!}\frac{%
d^{n}}{dt^{n}}\left[ \lim_{u^{2}\rightarrow 1}k_{\left\langle \mu _{1}\right. }\ldots k_{\left. \mu _{\ell
}\right\rangle }\int dK^{\prime }f_{0\mathbf{k}%
^{\prime }}\frac{\partial^{(\ell)} }{\partial u_{\mu _{1}}\cdots \partial u_{\mu
_{\ell }}}G_{t}\right] .
\end{equation}
We used that $\exp \left[
-tx/\left( 1-t\right) \right] /\left( 1-t\right) ^{\alpha +1}$ is the
generator of the associated Laguerre polynomial $L_n^{(\alpha)}(x)$ \cite{AbramowitzStegun}. For $\ell = 0$, the procedure introduced here becomes equivalent to the the one described in Ref.~\cite{Mullins:2022fbx}, where the scalar part of the spectrum of this linearized collision term was calculated.

It is convenient to remove the derivatives in the 4--velocity from the $dK^{\prime }$ integral. Since $f_{0\mathbf{k}^{\prime }}$ also depends on the $4$--velocity, this can be done by systematically applying the inverse chain rule to each derivative in $u^\mu$ and then calculating all possible permutations. Since the whole term is contracted with a symmetric tensor $k_{\left\langle \mu _{1}\right. }\ldots k_{\left. \mu
_{\ell }\right\rangle }$, all permutations can be trivially arranged. The integral $\mathcal{A}_{n\ell }$ then becomes 
\begin{eqnarray}
\mathcal{A}_{n\ell } &\equiv &g\frac{\left( 2\ell -1\right) !!}{\left(
-1\right) ^{\ell }E_{\mathbf{k}}^{2\ell }} \sum_{s=0}^{\ell }\frac{\beta^{s}}{s!\left( \ell -s\right) !}
 \lim_{t\rightarrow 0}\frac{1}{n!}\frac{d^{n}}{dt^{n}}\left[
\lim_{u^{2}\rightarrow 1}k_{\left\langle \mu _{1}\right. }\ldots k_{\left.
\mu _{\ell }\right\rangle }\frac{\partial ^{\left( \ell -s\right) }}{%
\partial u_{\mu _{1}}\cdots \partial u_{\mu _{\ell -s}}}\mathcal{I}_{t}^{\mu
_{\ell -s+1}\cdots \mu _{\ell }}\right] ,
\end{eqnarray}%
where we defined the $s$--th rank tensor%
\begin{equation}
\mathcal{I}_{t}^{\mu _{\ell -s+1}\cdots \mu _{\ell }}\equiv \int dK^{\prime
}f_{0\mathbf{k}^{\prime }}k^{\prime \mu _{\ell -s+1}}\ldots k^{\prime \mu
_{\ell }}G_{t}\text{,}
\end{equation}%
with all the derivatives of $f_{0\mathbf{k}^{\prime }}$
with respect to $u^{\mu }$ already being explicitly evaluated.

First, we note that $\mathcal{I}_{t}^{\mu _{\ell -s+1}\cdots \mu _{\ell }}$
depends only on two external 4-vectors, $u^{\mu }$ and $k^{\mu }$. However,
any term proportional to $u^{\mu _{i}}$, $i=0,\ldots,\ell$, will not contribute to $\mathcal{A}%
_{n\ell }$ since such terms are orthogonal to $k_{\left\langle \mu
_{1}\right. }\ldots k_{\left. \mu _{\ell }\right\rangle }$ and their
derivative with respect to $u^{\mu _{j}}$, $j=0,\ldots,\ell$, is constant $\partial u^{\mu
_{i}}/\partial u_{\mu _{j}}=g^{\mu _{i}\mu _{j}}$ and also vanishes once
contracted with $k_{\left\langle \mu _{1}\right. }\ldots k_{\left. \mu
_{\ell }\right\rangle }$ due to the tracelessness condition. Thus, the
tensor $k^{\prime \mu _{\ell -s+1}}\ldots k^{\prime \mu _{\ell }}$, in the
definition of $\mathcal{I}_{t}^{\mu _{\ell -s+1}\cdots \mu _{\ell }}$, can
be immediately replaced by its projection onto the 3-space orthogonal to $%
u^{\mu }$, $k^{\prime \left\langle \mu _{\ell -s+1}\right\rangle }\ldots
k^{\prime \left\langle \mu _{\ell }\right\rangle }$. We can further replace $%
k^{\prime \left\langle \mu _{\ell -s+1}\right\rangle }\ldots k^{\prime
\left\langle \mu _{\ell }\right\rangle }$ by its traceless and symmetric
projection $k^{\prime \left\langle \mu _{\ell -s+1}\right. }\ldots k^{\prime
\left. \mu _{\ell }\right\rangle }$. The term can be symmetrized because it
is contracted with a symmetric tensor and the trace can be removed since any
term proportional to the projection operator $\Delta ^{\mu _{i}\mu _{j}}$,
or its derivatives in $u^{\mu _{k}}$, $k=0,\ldots,\ell$, will vanish when contracted with $
k_{\left\langle \mu _{1}\right. }\ldots k_{\left. \mu _{\ell }\right\rangle }
$. Thus, in order to calculate $\mathcal{A}_{n\ell }$, we need only to
consider the irreducible projection of $\mathcal{I}_{t}^{\mu _{\ell
-s+1}\cdots \mu _{\ell }}$, 
\begin{eqnarray}
\mathcal{A}_{n\ell } &\equiv g&\sum_{s=0}^{\ell }\frac{ \beta^{s}}{s!\left( \ell -s\right) !}\frac{\left( 2\ell -1\right) !!}{\left(
-1\right)^{\ell }E_{\mathbf{k}}^{2\ell }}
 \lim_{t\rightarrow 0}\frac{1}{n!}\frac{d^{n}}{dt^{n}}\left[
\lim_{u^{2}\rightarrow 1}k_{\left\langle \mu _{1}\right. }\ldots k_{\left.
\mu _{\ell }\right\rangle }\frac{\partial ^{\left( \ell -s\right) }}{%
\partial u_{\mu _{1}}\cdots \partial u_{\mu _{\ell -s}}}\mathcal{I}%
_{t}^{\left\langle \mu _{\ell -s+1}\cdots \mu _{\ell }\right\rangle }\right]
.
\end{eqnarray}%
Finally, since $\mathcal{I}_{t}^{\mu _{\ell -s+1}\cdots \mu _{\ell }}$ can
only be constructed by combinations of $u^{\mu }$, $k^{\mu }$, and $g^{\mu
\nu }$, its irreducible projection must be of the following general form%
\begin{equation}
\mathcal{I}_{t}^{\left\langle \mu _{\ell -s+1}\cdots \mu _{\ell
}\right\rangle }=\mathcal{I}_{t}^{\left( s\right) }k^{\left\langle \mu
_{\ell -s+1}\right. }\ldots k^{\left. \mu _{\ell }\right\rangle },
\end{equation}
where we introduced yet another scalar integral, $\mathcal{I}_{t}^{\left(
s\right) }$%
\begin{equation}
\mathcal{I}_{t}^{\left( s\right) }=\frac{\left( 2s-1\right) !!}{\left(
-1\right) ^{s}s!E_{\mathbf{k}}^{2s}}\int dK^{\prime }f_{0\mathbf{k}^{\prime
}}k^{\left\langle \alpha _{1}\right. }\cdots k^{\left. \alpha
_{s}\right\rangle }k_{\left\langle \alpha _{1}\right. }^{\prime }\cdots
k_{\left. \alpha_{s}\right\rangle }^{\prime }G_{t}.
\end{equation}
This integral can be evaluated in terms of Kummer's confluent
hypergeometric function $M(a;b;x)$ \cite{AbramowitzStegun}, which gives
\begin{equation}
\mathcal{I}_{t}^{\left( s\right) }=\frac{\mathcal{M}t^{2s-\ell }}{\left(
u_{\alpha }u^{\alpha }\right) ^{s+1}\left( -\beta \right) ^{\ell }\left(
1-t\right) ^{\ell +s+1}}\frac{\left( s!\right) ^{2}}{\left( 2s+1\right) !}%
M\left( s+1;2s+2;\frac{t\beta E_{\mathbf{k}}}{t-1}\right) .
\label{eqdificil}
\end{equation}
The derivation of this expression is quite involved and is explained in detail in the Supplemental Material.

The derivatives in terms of the 4--velocity can be calculated using that 
\begin{eqnarray}
&&\lim_{u^{2}\rightarrow 1}k_{\left\langle \mu _{1}\right. }\ldots k_{\left.
\mu _{\ell }\right\rangle }\frac{\partial ^{\left( \ell -s\right) }}{%
\partial u_{\mu _{1}}\cdots \partial u_{\mu _{\ell -s}}}\left[ \frac{1}{%
\left( u_{\alpha }u^{\alpha }\right) ^{s+1}}M\left( s+1;2s+2;\frac{t}{t-1}%
\beta E_{\mathbf{k}}\right) k^{\left\langle \mu _{\ell -s+1}\right. }\ldots
k^{\left. \mu _{\ell }\right\rangle }\right]  \notag \\
&=&k_{\left\langle \mu _{1}\right. }\ldots k_{\left. \mu _{\ell
}\right\rangle }k^{\left\langle \mu _{1}\right. }\ldots k^{\left. \mu _{\ell
}\right\rangle }\left( \frac{t\beta }{t-1}\right) ^{\ell -s}\frac{\left(
2s+1\right) !\ell !}{s!\left( \ell +s+1\right) !}M\left( \ell +1;\ell +s+2;%
\frac{t\beta E_{\mathbf{k}}}{t-1}\right) ,
\end{eqnarray}%
where we note that all derivatives of $k^{\left\langle \mu _{\ell
-s+1}\right. }\ldots k^{\left. \mu _{\ell }\right\rangle }$ and $u_{\alpha
}u^{\alpha }$ with respect to $u^{\mu _{i}}$, $i\leq \ell -s$, are zero when
contracted with $k_{\left\langle \mu _{1}\right. }\ldots k_{\left. \mu
_{\ell }\right\rangle }$. We also used  the following property of $M\left(
a;b;x\right) $,%
\begin{equation}
\frac{d^{\ell -s}}{dx^{\ell -s}}M\left( s+1;2s+2;x\right) =\frac{\left(
2s+1\right) !\ell !}{s!\left( \ell +s+1\right) !}M\left( \ell +1;\ell
+s+2;x\right) .
\end{equation}%
With this, we find the following expression for the integral $\mathcal{A}_{n\ell }$,
\begin{equation}
\mathcal{A}_{n\ell }=g\mathcal{M}\sum_{s=0}^{\ell }\frac{\left( \ell
!\right) ^{2}\left( -1\right) ^{s}}{\left( \ell -s\right) !\left( \ell
+s+1\right) !}\lim_{t\rightarrow 0}\frac{1}{n!}\frac{d^{n}}{dt^{n}}\left[ 
\frac{t^{s}}{\left( 1-t\right) ^{2\ell +1}}M\left( \ell +1;\ell +s+2;\frac{%
t\beta E_{\mathbf{k}}}{t-1}\right) \right] .
\end{equation}

We now use the integral expression of $M( a;b;z)$ \cite{AbramowitzStegun}
\begin{equation}
M\left( a;b;z\right) =\frac{\Gamma \left( b\right) }{\Gamma \left( a\right)
\Gamma \left( b-a\right) }\int_{0}^{1}dx\left. {}\right. \exp \left(
zx\right) x^{a-1}\left( 1-x\right) ^{b-a-1},
\end{equation}%
and the fact that $\exp \left[ -tx/\left( 1-t\right) \right] /\left(
1-t\right) ^{2\ell +1}$ is the generator of the associate Laguerre polynomial, $L_{n}^{\left( 2\ell \right)}\left( x\right)$, to express $\mathcal{A}_{n\ell }$ in the following form
\begin{equation}
\mathcal{A}_{n\ell }=g\mathcal{M}\sum_{s=0}^{\ell }\sum_{i=0}^{\infty }\frac{%
\ell !\left( -1\right) ^{s}}{\left( \ell -s\right) !s!}\lim_{t\rightarrow 0}%
\frac{1}{n!}\frac{d^{n}}{dt^{n}}\left[ t^{i+s}\int_{0}^{1}dx\left. {}\right.
x^{\ell }\left( 1-x\right) ^{s}L_{i}^{\left( 2\ell \right) }\left( \beta E_{%
\mathbf{k}}x\right) \right] ,
\end{equation}
and explicitly calculate its derivatives with respect to $t$,
\begin{equation}
\mathcal{A}_{n\ell }=g\mathcal{M}\sum_{s=0}^{n}\frac{\ell !\left( -1\right)
^{s}}{\left( \ell -s\right) !s!}\int_{0}^{1}dx\left. {}\right. x^{\ell
}\left( 1-x\right) ^{s}L_{n-s}^{\left( 2\ell \right) }\left( \beta E_{%
\mathbf{k}}x\right) .
\end{equation}
The integral in $x$ can be evaluated by expressing the associate Laguerre polynomial as a series
in powers of $\beta E_{\mathbf{k}}x$. All the remaining summations can be explicitly evaluated, and using that 
\begin{equation}
\sum_{s=0}^{i}\frac{\ell !\left( -1\right) ^{s+i}}{\left( \ell +s-i\right) !}%
\frac{\left( n+2\ell +s-i\right) !}{s!\left( n+\ell +1-s\right) !}=\frac{1}{%
\left( n+\ell +1\right) \left( 2\ell +n+1-i\right) }\frac{\left( n+2\ell
+1\right) !}{i!\left( \ell +n-i\right) !},
\end{equation}
we confirm that $\mathcal{A}_{n\ell }$ is proportional to a Laguerre polynomial, i.e., 
\begin{equation}
\mathcal{A}_{n\ell } =\frac{g\mathcal{M}}{n+\ell +1}L_{n}^{\left( 2\ell +1\right) }\left( \beta
E_{\mathbf{k}}\right) .
\end{equation}

Finally, combining this calculation with the result derived for the loss term \eqref{loss}, we obtain
\begin{equation}
\hat{L}L_{n\mathbf{k}}^{\left( 2\ell +1\right) }k^{\left\langle \mu
_{1}\right. }\ldots k^{\left. \mu _{\ell }\right\rangle }=-\frac{g\mathcal{M}%
}{2}\left[ \frac{n+\ell -1}{n+\ell +1}+\delta _{\ell 0}\delta _{n0}\right]
L_{n\mathbf{k}}^{\left( 2\ell +1\right) }k^{\left\langle \mu _{1}\right.
}\ldots k^{\left. \mu _{\ell }\right\rangle },
\label{mainresult}
\end{equation}
and we conclude, as initially stated, that $L_{n\mathbf{k}}^{\left( 2\ell
+1\right) }k^{\left\langle \mu _{1}\right. }\ldots k^{\left. \mu _{\ell
}\right\rangle }$ are eigenfunctions of $\hat{L}$, with the corresponding discrete set of eigenvalues $\chi
_{n\ell }$%
\begin{equation}
\chi _{n\ell }=-\frac{g\mathcal{M}}{2}\left( \frac{n+\ell -1}{n+\ell +1}%
+\delta _{\ell 0}\delta _{n0}\right) ,
\label{eigenvalues}
\end{equation}
with $n,\ell$ being non-negative integers\footnote{We note that one recovers the results for the eigenvalues derived in \cite{Mullins:2022fbx} by taking $\ell=0$.}.
As expected, we find that the eigenvalues vanish when the eigenfunctions correspond to quantities that are conserved in elastic collisions, i.e., for $(n,\ell)=(0,0),(1,0),(0,1)$. The remaining nonzero eigenvalues, which determine all the non-hydrodynamic modes \cite{forster1995hydrodynamic} of this theory and thus dictate the rates of relaxation of the gas to its equilibrium distribution, are negative in accordance with the stability property of the global equilibrium state \cite{cercignani:90mathematical}. Furthermore, it is interesting to note that the eigenvalue spectrum is bounded from below, i.e., $\chi_{n\ell} \geq  -g\mathcal{M}/2$. This is in sharp contrast with the case of Maxwell molecules in a non-relativistic gas where the eigenvalues grow with the fourth root of an integer \cite{cercignani:90mathematical}. 

Our results provide the first expression for the full set of eigenvalues and eigenvectors of the relativistic linearized Boltzmann equation. Even in the nonrelativistic case, where the eigenfunctions of the linearized Boltzmann equation have been determined for a gas of  Maxwell molecules \cite{ChangUhlenbeck,cercignani:90mathematical}, simple analytical expressions for the eigenvalues were not found. Naturally, once the spectral decomposition of the operator is determined, a plethora of new calculations and applications can be carried out, i.e., the transport properties of the system can be easily determined, the hydrodynamic limit can be explicitly evaluated, the convergence or not of the hydrodynamic series can be analyzed, and the linearized Boltzmann equation can be straightforwardly solved using the method of moments.\\ 

\noindent
{\it \textbf{Transport coefficients}} -- We begin to explore the consequences of our results by providing exact expressions for the shear and particle diffusion transport coefficients in $\lambda \phi^4$ theory. This will be done employing the traditional Chapman-Enskog theory –- a perturbative solution of the Boltzmann equation based in a derivative expansion \cite{Denicol:2021,degroot}. The zeroth-order solution of this
expansion is the local equilibrium solution itself, i.e., $\phi_{\mathbf{k}}=0$. The first-order solution for $\phi_{\mathbf{k}}$ is more complicated and must be obtained by inverting the following equation,
\begin{equation}
 \frac{1}{4}L^{(3)}_{1\mathbf{k}}k_{\langle\mu\rangle}\nabla ^{\mu }\alpha -\beta 
k_{\langle \mu }k_{\mu \rangle}
\sigma^{\mu \nu } 
=\hat{L}\phi_{\mathbf{k}} . \label{CE}
\end{equation}%
Above, we have already imposed the massless and classical limits and, for the sake of convenience, expressed the left-hand side of the equation in terms of associate Laguerre polynomials. We further defined $\nabla^\mu \equiv \partial^{\langle \mu \rangle}$ and $\sigma^{\mu \nu} \equiv \partial^{\langle \mu} u^{\nu \rangle}$.

The general solution for $\phi_{\mathbf{k}}$ is formally given by
\begin{equation}
\phi_{\mathbf{k}} = \phi^{\mathrm{hom}}_{\mathbf{k}} + \hat{L}^{-1}\left(\frac{1}{4}L^{(3)}_{1\mathbf{k}}k_{\langle\mu\rangle}\nabla ^{\mu }\alpha -\beta 
k_{\langle \mu }k_{\mu \rangle}
\sigma^{\mu \nu } \right),
 \label{CEsol}
\end{equation}%
where $\phi^{\mathrm{hom}}_{\mathbf{k}}=a + b_\mu k^\mu$ is the homogeneous solution, with the free parameters $a$ and $b_\mu$ being determined by matching conditions \cite{Denicol:2021}. Since we have obtained the eigenvalues and eigenfunctions of the linear operator $\hat{L}$, we can explicitly evaluate the solution above,
\begin{equation}
\phi_{\mathbf{k}} = a + b_\mu k^\mu + \frac{1}{4\chi_{11}}L^{(3)}_{1\mathbf{k}}k_{\langle\mu\rangle}\nabla ^{\mu }\alpha -\frac{\beta}{\chi_{02}} 
k_{\langle \mu }k_{\mu \rangle}
\sigma^{\mu \nu },
 \label{CEsol}
\end{equation}
with explicit expressions for $\chi_{11}$ and $\chi_{02}$ being given in \eqref{eigenvalues}. Here, we employ Landau matching conditions \cite{LandauLifshitzFluids}, which impose that the following moments of $f_{0\mathbf{k}}\phi_{\mathbf{k}}$ vanish,   
\begin{eqnarray}
 \int dK E_{\mathbf{k}} f_{0\mathbf{k}}\phi_{\mathbf{k}} = 0,
 \int dK E_{\mathbf{k}} k^\mu f_{0\mathbf{k}}\phi_{\mathbf{k}} = 0.
 \label{CEsol}
\end{eqnarray}
Then, one determines the free parameters of the homogeneous solution to be $a=0$ and $b^\mu=\nabla^\mu \alpha / (4\chi_{11})$. 

The particle diffusion 4-current, $n^{\mu}$, and shear stress tensor, $\pi^{\mu\nu}$, are obtained by replacing this solution for $\phi_{\mathbf{k}}$ into the definitions of these dissipative currents,  
\begin{eqnarray}
n^\mu \equiv \int dK k^{\langle \mu \rangle} f_{0\mathbf{k}}\phi_{\mathbf{k}} , \qquad 
\pi^{\mu\nu} \equiv \int dK k^{\langle \mu}k^{\nu \rangle} f_{0\mathbf{k}} \phi_{\mathbf{k}} .   \label{definitions}
\end{eqnarray}
This procedure leads to the relativistic Navier-Stokes equations where, $n^\mu=\kappa_n \nabla^{\mu} \alpha$ and $\pi^{\mu\nu}=2\eta \sigma^{\mu\nu}$, with $\kappa_n$ being the particle diffusion coefficient, and $\eta$ the shear viscosity. The exact expressions for these coefficients can now be obtained using the orthogonality conditions satisfied by the irreducible tensors and Laguerre polynomials, and are given by,
\begin{eqnarray}
\kappa_n = \frac{3}{g\beta^2} , \qquad
\eta =  \frac{48}{g\beta^3}.  
\label{transport}
\end{eqnarray}
As expected of a conformal system (in the massless limit), $\kappa_n$ goes with $\sim T^2$ while $\eta \sim T^3$. Exact expressions for these transport coefficients, for any type of interactions, do not exist in the literature and are always obtained by numerically inverting the linearized collision operator. 

\noindent
{\it \textbf{Relaxation time approximation}} -- Once the spectral decomposition of $\hat{L}$ is known, we can determine approximate expressions for the linearized collision term in the form usually associated with the relaxation time approximation. First, we expand $\phi_{\mathbf{k}}$ in the complete basis of polynomials and irreducible tensors that constitute the eigenfunctions of $\hat{L}$:  
\begin{equation}
\phi _{\mathbf{k}}=\sum_{n,\ell =0}^{\infty }c_{n}^{\mu _{1}\cdots \mu
_{\ell }}L_{n\mathbf{k}}^{\left( 2\ell +1\right) }
k_{\left\langle \mu _{1}\right. }\ldots k_{\left. \mu _{\ell }\right\rangle },
\label{expansion}
\end{equation}
where the expansion coefficients $c_{n}^{\mu _{1}\cdots \mu
_{\ell }}$ can be determined using the orthogonality conditions satisfied by the basis elements \cite{Denicol:2021} and can be expressed as the following integrals of $\phi_{\mathbf{k}}$, 
\begin{eqnarray}
\int dKf_{0\mathbf{k}}\phi _{\mathbf{k}}k^{\left\langle \mu _{1}\right.
}\ldots k^{\left. \mu _{\ell }\right\rangle }L_{n}^{\left( 2\ell +1\right) }
=\left(-1\right)^{\ell } \frac{\left( n+2\ell +1\right)! \ell! \mathcal{M}}
{n! \left(2\ell +1\right)!!  \beta ^{2\ell}} c_{n}^{\mu _{1}\cdots \mu _{\ell }}.
\end{eqnarray}
The linearized collision term appearing in the Boltzmann equation is $\hat{L}\phi _{\mathbf{k}}$ and, using expansion \eqref{expansion} and our main result \eqref{mainresult}, it can be written as
\begin{eqnarray}
\hat{L}\phi _{\mathbf{k}} 
=-\frac{g \mathcal{M}}{2}\left[ \phi _{\mathbf{k}%
}-\sum_{n,\ell =0}^{\infty }c_{n}^{\mu _{1}\cdots \mu _{\ell }}\left( \frac{2%
}{n+\ell +1}-\delta _{\ell 0}\delta _{n0}\right) L_{n\mathbf{k}}^{\left(
2\ell +1\right) }k_{\left\langle \mu _{1}\right. }\ldots k_{\left. \mu
_{\ell }\right\rangle }\right] ,
\label{RTA}
\end{eqnarray}%
where, above, we isolated the contribution proportional to $\phi_{\mathbf{k}}$.

We note that in the exact result in \eqref{RTA} 
the terms in the sum are suppressed as $\sim 1/(n+\ell)$. The relaxation time approximation corresponds to simply neglecting such terms. If we only retain the first contribution on the right hand side of \eqref{RTA}, we obtain the well-known Anderson-Witting approximation \cite{ANDERSON1974466} for the linearized collision term, $\hat{L}\phi _{\mathbf{k}} \approx -(E_{\mathbf{k}}/\tau_{R}) \phi _{\mathbf{k}}$, with the relaxation time being identified as $\tau_{R}=2E_{\mathbf{k}}/(g \mathcal{M})$. However, this approximation was shown to be flawed in Ref.~\cite{Rocha:2021zcw}, since it does not maintain the fundamental properties of $\hat{L}$ that emerge from the conservation laws in microscopic collisions. These properties can be preserved by keeping the terms corresponding to eigenfunctions with vanishing eigenvalues, that is
\begin{eqnarray}
\hat{L}\phi _{\mathbf{k}} &\approx &-\frac{g \mathcal{M}}{2}\left[ \phi _{\mathbf{k}}-c_{0}-c_{1}L_{1\mathbf{k}%
}^{\left( 1\right) }-c_{0}^{\mu }k_{\left\langle \mu \right\rangle }\right]
=-\frac{E_{\mathbf{k}}}{\tau _{R}}\left[ \phi _{\mathbf{k}}-%
\frac{\left\langle \phi _{\mathbf{k}%
}\right\rangle _{0}}{\left\langle 1
\right\rangle _{0}}-L_{1\mathbf{k}}^{\left( 1\right) }\frac{\left\langle \phi _{\mathbf{k}}L_{1}^{\left(
1\right) }\right\rangle _{0}}{\left\langle  L_{1}^{\left( 1\right) }L_{1}^{\left( 1\right) }\right\rangle
_{0}}-k_{\left\langle \mu \right\rangle }\frac{\left\langle \phi _{\mathbf{k}}k^{\left\langle \mu
\right\rangle }\right\rangle _{0}}{\frac{1}{3}\left\langle k_{\left\langle \mu \right\rangle }k^{\left\langle \mu
\right\rangle }\right\rangle _{0}}\right],
\label{rtafinal}
\end{eqnarray}
where we used the notation $\left\langle\cdots \right\rangle_{0} \equiv \int dK \left( E_{\mathbf{k}}/\tau _{R}\right)f_{0\mathbf{k}}\left(\cdots \right)$, and the relaxation time is again identified as $\tau_{R}=2E_{\mathbf{k}}/(g \mathcal{M})$. We then see that \eqref{rtafinal} assumes exactly the form of the new relaxation time approximation proposed in Ref.~\cite{Rocha:2021zcw}. However, in this paper we were able to derive this approximate form of $\hat{L}$ \textit{and} an exact expression for the relaxation time. Usually, the relaxation time is left as a free parameter of this \textit{Ansatz} and is determined so that the shear viscosity is well reproduced by the model. This procedure will not lead to the correct expression for the relaxation time derived above.

\noindent
{\it \textbf{Conclusions}} -- We analytically solved the eigenvalue problem defined by the linearized Boltzmann collision operator of massless scalar $\lambda\phi^4$ theory in the high-temperature regime. This is the first time that both the eigenfunctions and the eigenvalue spectrum of a linearized Boltzmann collision operator are fully determined in closed form. Our results can be useful in the investigation of a variety of problems involving the emergence of relativistic hydrodynamic behavior from kinetic theory. Immediate applications that can be pursued include the determination of the properties of the hydrodynamic expansion \cite{Heller:2013fn,Heller:2015dha,Buchel:2016cbj,Denicol:2016bjh,Heller:2016rtz,Strickland:2017kux,Grozdanov:2019kge,Denicol:2019lio,Almaalol:2020rnu,Heller:2021yjh}, and the computation of the transport coefficients and the effects of higher-order moments on the derivation of causal theories of relativistic hydrodynamics  \cite{Denicol:2012cn,Bemfica:2017wps,Bemfica:2019knx,Kovtun:2019hdm,Hoult:2020eho,Bemfica:2020zjp,Rocha:2022ind}. The extension of our results to take into account the bosonic nature of the excitations, the inclusion of a nonzero mass, or other type of interactions (e.g., including gauge and fermion fields) should also be actively pursued. A more challenging problem involves going beyond the linearized approximation, considering the full nonlinear collision operator. We hope our results can shed light on some of these problems and pave the way for a new understanding of relativistic gases.

\noindent
\section{Acknowledgements} G.~S.~D.~acknowledge Conselho Nacional de Desenvolvimento Cient\'ifico e Tecnol\'ogico (CNPq) and Funda\c c\~ao Carlos Chagas Filho de Amparo \`a Pesquisa do Estado do Rio de Janeiro (FAPERJ) for financial support. J.~N.~is partially supported by the U.S.~Department of Energy, Office of Science, Office for Nuclear Physics under Award No. DE-SC0021301. The authors thank Funda\c c\~ao de Amparo \`a Pesquisa do Estado de S\~ao Paulo (FAPESP), grant number 2017/05685-2, for support.

\bibliography{References}{}
\bibliographystyle{bibstyl}

\newpage

\appendix

\supplemental{SUPPLEMENTAL MATERIAL}
\section{Generating function and $\mathcal{I}_{t}^{\left( s\right) }$}

In this Supplemental Material, we calculate the generating function $G_{t}$
and the integral $\mathcal{I}_{t}^{\left( s\right) }$. The generating
function is a Lorentz scalar that depends solely on $u^{\mu }$ and the total
4-momentum $P_{T}^{\mu }=k^{\mu }+k^{\prime \mu }$. Therefore, it must be a
function of all the possible Lorentz scalars that can be constructed from
these 4-vectors: these are $u^{\mu }u_{\mu }\equiv u^{2}$, $P_{T}^{\mu
}P_{T\mu }\equiv s$, and $\hat{P}_{T\mu }\hat{u}^{\mu }$, with $\hat{P}%
_{T}^{\mu }=P_{T}^{\mu }/\sqrt{s}$ and $\hat{u}^{\mu }/u$ being the
normalized versions of the corresponding 4-vectors. This implies that $G_{t}$
is invariant under the exchange $\hat{u}^{\mu }\leftrightarrow \hat{P}%
_{T}^{\mu }$ and, without loss of generality, we can rewrite this integral as%
\begin{equation}
G_{t}\equiv \frac{\left( t\beta \right) ^{2}}{\left[ -t\beta \left(
1-t\right) \right] ^{\ell +2}}\int dPdP^{\prime }\left. {}\right. \left(
2\pi \right) ^{5}\delta ^{4}\left( \sqrt{s}\hat{u}-p-p^{\prime }\right) \exp
\left( -\frac{t}{1-t}\beta up_{\mu }\hat{P}_{T}^{\mu }\right) .
\end{equation}%
In the local rest frame, $\hat{u}^{\mu }=\left( 1,0,0,0\right) $, the delta
function is considerably simplified and can be trivially evaluated, leaving
only the integral over $\mathbf{\hat{p}}=\mathbf{p}/p$,%
\begin{equation}
G_{t}\equiv \frac{\left( t\beta \right) ^{2}}{\left[ -t\beta \left(
1-t\right) \right] ^{\ell +2}}\int \frac{d\mathbf{\hat{p}}}{4\pi }\left.
{}\right. \exp \left[ -\frac{tu\beta k}{2\left( 1-t\right) }\left( 1-\mathbf{%
\hat{p}}\cdot \mathbf{\hat{k}}\right) \right] \exp \left[ -\frac{tu\beta
k^{\prime }}{2\left( 1-t\right) }\left( 1-\mathbf{\hat{p}}\cdot \mathbf{\hat{%
k}}^{\prime }\right) \right],
\end{equation}%
where we used the notation $p=\left\vert \mathbf{p}\right\vert $.

Using the identities%
\begin{eqnarray}
\delta \left( x-y\right)  &=&\sum_{q=0}^{\infty }\left( q+\frac{1}{2}\right)
P_{q}\left( x\right) P_{q}\left( y\right) , \\
P_{q}\left( \mathbf{\hat{k}}\cdot \mathbf{n}_{\ell }\right)  &=&\frac{4\pi }{%
2q+1}\sum_{m_{q}=-q}^{q}Y_{q}^{m_{q}}\left( \Theta _{\ell },\Phi _{\ell
}\right) Y_{q}^{\ast m_{q}}\left( \Theta _{k},\Phi _{k}\right),
\end{eqnarray}%
where $P_{q}\left( x\right) $ is the $q$--th order Legendre polynomial and $%
Y_{q}^{m_{q}}$ are the spherical harmonics, and the orthogonality relation
satisfied by the spherical harmonics, one can show that%
\begin{eqnarray}
G_{t} &=&\frac{\left( t\beta \right) ^{2}}{\left[ -t\beta \left( 1-t\right) %
\right] ^{\ell +2}}\sum_{q=0}^{\infty }\frac{2q+1}{4}P_{q}\left( \mathbf{%
\hat{k}\cdot \hat{k}}^{\prime }\right)  \notag\\
&&\times \int dxdy\exp \left[ -\frac{tu\beta k\left( 1-x\right) }{2\left(
1-t\right) }\right] \exp \left[ -\frac{tu\beta k^{\prime }\left( 1-y\right) 
}{2\left( 1-t\right) }\right] \text{ }P_{q}\left( x\right) P_{q}\left(
y\right) .
\end{eqnarray}%
We note that this results is expressed in the local rest frame of the system.

Replacing this result into the expression for $\mathcal{I}_{t}^{\left(
s\right) }$ and using that, in the local rest frame of the fluid \cite{Denicol:2021},
\begin{equation}
k^{\left\langle \mu _{1}\right. }\cdots k^{\left. \mu _{\ell }\right\rangle
}p_{\left\langle \mu _{1}\right. }\cdots p_{\left. \mu _{\ell }\right\rangle
}=\frac{\left( -2kp\right) ^{\ell }\left( \ell !\right) ^{2}}{\left( 2\ell
\right) !}P_{\ell }\left( \mathbf{\hat{k}}\cdot \mathbf{\hat{p}}\right) ,
\end{equation}%
we obtain the expression,%
\begin{eqnarray}
\mathcal{I}_{t}^{\left( s\right) } &=&\frac{\left( t\beta \right) ^{2}}{%
\left[ -t\beta \left( 1-t\right) \right] ^{\ell +2}}\sum_{q=0}^{\infty }%
\frac{2q+1}{4u^{2s}}\int dK^{\prime }f_{0\mathbf{k}^{\prime }}\frac{%
k^{\prime s}}{k^{s}}P_{s}\left( \mathbf{\hat{k}}\cdot \mathbf{\hat{k}}%
^{\prime }\right) P_{q}\left( \mathbf{\hat{k}\cdot \hat{k}}^{\prime }\right) \notag
\\
&&\times \int dxdy\exp \left[ -\frac{tu\beta k\left( 1-x\right) }{2\left(
1-t\right) }\right] \exp \left[ -\frac{tu\beta k^{\prime }\left( 1-y\right) 
}{2\left( 1-t\right) }\right] \text{ }P_{q}\left( x\right) P_{q}\left(
y\right) .
\end{eqnarray}%
The Legendre polynomials carry all the angular dependence in $\mathbf{k}%
^{\prime }$ and the orthogonality relation satisfied by these special
functions will guarantee that only terms with $q=s$ can survive. Thus, we
have that 
\begin{eqnarray}
\mathcal{I}_{t}^{\left( s\right) } &=&\frac{\left( t\beta \right) ^{2}}{%
4u^{2s}\left[ -t\beta \left( 1-t\right) \right] ^{\ell +2}}\int dK^{\prime
}f_{0\mathbf{k}^{\prime }}\frac{k^{\prime s}}{k^{s}}\int dx\exp \left[ -%
\frac{tu\beta k\left( 1-x\right) }{2\left( 1-t\right) }\right] \text{ }%
P_{s}\left( x\right)  \notag\\
&&\times \int dy\exp \left[ -\frac{tu\beta k^{\prime }\left( 1-y\right) }{%
2\left( 1-t\right) }\right] \text{ }P_{s}\left( y\right) .
\end{eqnarray}%
We proceed by performing the integral in $k^{\prime }$, leading to 
\begin{eqnarray}
\mathcal{I}_{t}^{\left( s\right) } &=&\mathcal{M}\frac{\left( t\beta \right)
^{2}}{4u^{2s+2}\left[ -t\beta \left( 1-t\right) \right] ^{\ell +2}}\int
dx\exp \left[ -\frac{tu\beta k\left( 1-x\right) }{2\left( 1-t\right) }\right]
\text{ }P_{s}\left( x\right)  \notag\\
&&\times \int dy\text{ }P_{s}\left( y\right) \frac{\left( s+1\right) !}{%
\left( \beta uk\right) ^{s}\left[ 1+\frac{t\left( 1-y\right) }{2\left(
1-t\right) }\right] ^{s+2}}.
\end{eqnarray}
The integrals in $x$ and $y$ are evaluated by expanding the integrand in
powers of $\left( 1-y\right) \ $and $\left( 1-x\right) $ via a Taylor series and  using the result \cite{AbramowitzStegun},
\begin{equation}
\int_{-1}^{1}dy\left( 1-y\right) ^{a+s}P_{s}\left( y\right) =\frac{\left(
-1\right) ^{s}2^{a+s+1}\left[ \left( a+s\right) !\right] ^{2}}{\left(
a+2s+1\right) !a!}.
\end{equation}%
One then obtains the following expression 
\begin{equation}
\mathcal{I}_{t}^{\left( s\right) }=\frac{\mathcal{M}t^{2s-\ell }}{%
u^{2s+2}\left( -\beta \right) ^{\ell }\left( 1-t\right) ^{\ell +s+1}}\frac{%
\left( s!\right) ^{2}}{\left( 2s+1\right) !}M\left( s+1;2s+2;\frac{t}{t-1}%
u\beta k\right) ,
\end{equation}%
where we used that%
\begin{eqnarray}
\sum_{i=0}^{\infty }\left( -\frac{t}{1-t}\right) ^{i+s}\frac{\left(
i+s\right) !}{i!} &=&s!\left( 1-t\right) \left( -t\right) ^{s}, \\
\sum_{j=0}^{\infty }\frac{\left( j+s\right) !}{\left( j+s+s+1\right) !}\frac{%
a^{j}}{j!} &=&\frac{s!}{\left( 2s+1\right) !}M\left( s+1;2s+2;a\right),
\end{eqnarray}%
with $M\left( a;b;z\right) $ being Kummer's (confluent hypergeometric) function. In a general reference frame, this can be written as
\begin{equation}
\mathcal{I}_{t}^{\left( s\right) }=\frac{\mathcal{M}t^{2s-\ell }}{\left(
u_{\alpha }u^{\alpha }\right) ^{s+1}\left( -\beta \right) ^{\ell }\left(
1-t\right) ^{\ell +s+1}}\frac{\left( s!\right) ^{2}}{\left( 2s+1\right) !}%
M\left( s+1;2s+2;\frac{t\beta E_{\mathbf{k}}}{t-1}\right),
\end{equation}
where we use the notation $E_{\mathbf{k}}=u_\mu k^\mu$. This is Eq.\ \eqref{eqdificil} of the main text.

\end{document}